\documentclass[twocolumn,twocolappendix]{aastex631}

\newcommand{\lFig}[1]{{\label{fig:#1}}}

\newcommand{\lTab}[1]{{\label{tab:#1}}}
\newcommand{\FIGFF}[2]{{\ref{fig:#2}{#1}}}

\newcommand{\FIG}[2]{{Fig.~\FIGFF{#1}{#2}}}
\newcommand{\Fig}[1]{{\FIG{}{#1}}}

\newcommand{\Msun}{\ensuremath{\mathrm{M}_\odot}}

\newcommand{\Tab}[1]{{Table \ref{tab:#1}}}

\begin{document}

\title[Stellar merger structure]{The structure and evolution of a high-mass stellar merger in the Hertszprung gap}

\author{Rachel A. Patton}
\affiliation{Department of Astronomy, The Ohio State University, 140 West 18th Ave, Columbus, OH 43210, USA}
\affiliation{Center for Cosmology and AstroParticle Physics, The Ohio State University,191 West Woodruff Avenue, Columbus, OH 43210, USA}
\affiliation{Department of Physics and Astronomy and Pittsburgh Particle Physics, Astrophysics and Cosmology Center (PITT PACC), University
of Pittsburgh, 3941 O’Hara Street, Pittsburgh, PA 15260, USA}

\author[0000-0002-7549-7766]{Marc. H. Pinsonneault}
\affiliation{Department of Astronomy, The Ohio State University, 140 West 18th Ave, Columbus, OH 43210, USA}
\affiliation{Center for Cosmology and AstroParticle Physics, The Ohio State University,191 West Woodruff Avenue, Columbus, OH 43210, USA}

\author[0000-0003-2377-9574 ]{Todd A. Thompson}
\affiliation{Department of Astronomy, The Ohio State University, 140 West 18th Ave, Columbus, OH 43210, USA}
\affiliation{Center for Cosmology and AstroParticle Physics, The Ohio State University,191 West Woodruff Avenue, Columbus, OH 43210, USA}
\affiliation{Department of Physics, The Ohio State University, 191 W Woodruff Ave, Columbus, OH 43210, USA}




\begin{abstract}

Post-main-sequence binary mergers are a common evolutionary pathway for massive stars, but the effects of merging on the long-term structure and evolution of the resulting star are a matter of active debate. Furthermore, the way in which merger products are modeled in 1D is not uniform. We present the evolution of an 11 \Msun~and 6.6 \Msun~binary on an 11 day orbit, that merges while the primary is crossing the Hertzsprung gap. We construct the merger product either by rapidly accreting the secondary onto the surface of the primary or by injecting material from the secondary deeper into the primary via entropy sorting. We then evolve them to carbon ignition, comparing their interior structures at this stage. We find that all merger products experience an extended blue supergiant phase and have undermassive helium cores and low carbon mass fractions compared to single and stripped stars. However, the evolution of central density, temperature, and composition in the entropy-sorted model is distinct from those of the rapid-accretion models.

\end{abstract}


\section{Introduction}
The standard theory of stellar structure is powerful and has been remarkably successful. However, single-star evolution is typically treated as the norm. In reality, many stars interact with their companions, which can induce drastic structural changes \citep[for recent examples see][]{Pat20,Rui21,Lap21,Sch21,Ren23,Gil25}. This is particularly important for massive stars for two reasons. First, massive stars live primarily in binaries and higher-order systems \citep{Bla61,Abt83,Egg08,San13,Moe17,Off23}. Many of these systems will interact. For example, an observational study of O-star binaries in the Tarantula Nebula suggests that $\sim$ 50\% of the binaries will interact \citep{San13}. This rate rises to $\sim$ 70\% \citep{San12} for a subset of the Galactic O-star binary population \citep{San12}.

Second, whether a massive star produces a successful supernova or collapses into a stellar mass back hole at the end of its life, as well as the properties of their respective associated transients, depends sensitively on the structure of the star's core at the time of core-collapse \citep[e.g.,][and references therein]{Oco11,Ugl12,Pej15,Suk14,Ert16,Mul16,Suk16,Suk18,Var18,Suk20,Pat20,Tak23,Bur24}. The final structure at collapse depends sensitively, on the core structure at the time of carbon ignition \citep{Tim96,Bro00,Bro01,Suk14,Suk16,Suk20,Pat20,Lap25}, though binary interaction after carbon ignition could also change the core of the star prior to core-collapse, and the carbon-oxygen (CO) core structure at carbon ignition is set by the prior evolution of the star. 

There are a multitude of ways in which close binaries can interact, depending on the initial period $P$, the masses of the primary $M_1$ and secondary $M_2$, and the mass ratio ($q \equiv \frac{M_2}{M_1}$) of the stars in the system. Interactions are divided into three categories: case A occurs on the main sequence (orbital periods of a few days); case B occurs when the donor star is crossing the Hertzsprung gap (orbital periods of a few days to a few hundred days); case C occurs during core helium burning and beyond (orbital periods of a few hundred days to a few years) \citep{Kip67, Lau69, Pac71}. The timing of when stars begin to interact depends most strongly on the initial orbital period: case A interactions occur at shorter initial periods, whereas case C occur at much longer initial periods \citep{Pac71,Pod92a,Pol94}. Different histories give rise to different structures, ultimately impacting the final fates of these stars. For example, case A mergers produce stars whose subsequent structure, evolution, and final fates are comparable, though not equivalent, to an initially single star of comparable mass \citep{Gle13, Hen25}. Alternatively, when stars have their envelopes completely stripped via case B mass transfer, the extent of the convective helium cores recede \citep{Woo19,Lap21}. This changes the relative rates of the triple-alpha process and $\alpha$-capture onto $^{12}$C such that the amount of carbon in the core increases. If higher mass cores have more carbon, they may be able to burn carbon convectively instead of radiatively, which ultimately leads to cores which are easier to explode \citep[e.g.,][]{Tim96,Pat20,Lap21,Sch21,Pat22}.

However, the landscape of possible binary interactions is complex, and the long-term structural and evolutionary impacts of these interactions are poorly understood. Yet, binary interaction, and merging in particular, is very common. Roughly a quarter to third of massive stars will merge with a companion during their lifetime \citep{San12,Ren19,Zap19}. For this paper, we focus on case B mergers. Case B mass transfer is the most common type of mass transfer, and a non-negligible fraction of case B mass transfer systems should lead to a merger \citep{Pod10,Schu24}. Case B mergers arise from two different formation channels depending on the mass ratio of the binary and the response of the donor star to mass transfer \citep{Pod92a,Pol94,Schu24}. If the donor star responds to mass transfer by expanding (in the fully adiabatic, polytropic limit, this occurs when the star has a convective envelope), mass transfer can become unstable \footnote{The Roche lobe of both the accretor and the donor can also grow or shrink in response to mass transfer. In principle, unstable mass transfer could also occur if the donor's Roche lobe shrinks faster than its radius.}. The accretor cannot thermally adjust to the mass it is gaining faster than the mass transfer rate, causing it to fill its Roche lobe. This leads to a common envelope. The accretor and the core of the donor orbit one another in this common envelope while drag from the envelope causes the orbits to decay. If the orbits decay faster than the envelope can be ejected, the accretor and the core of the donor will merge. This typically occurs for smaller mass ratios \citep[e.g.][]{Zap19}. The physics governing this entire process remains highly uncertain \citep{Iva13,Iva20,Rop23}.

If the donor responds by contracting (donors with radiative envelopes in the fully adiabatic, polytropic limit) mass transfer can remain stable, but the binary can still come into contact as the accretor fills its Roche lobe \citep{Pod92a,Pol94,Jus14}. The range in mass ratio over which this occurs is uncertain and sensitive to model assumptions. The models of \citet{Pol94} and \citet{Jus14} put the limits between mass ratios of $\sim$ 0.5 and 0.8 for conservative mass transfer scenarios, whereas the models from \citet{Hen24} show a strong dependence of the range of mass ratios over which contact is achievable depending on how conservative mass transfer is. Once in contact, the continued expansion of the donor as it moves across the Hertzsprung gap can lead to the stars merging. This process is accelerated if material overflows the outer Lagrange point ($L_2$) carrying away angular momentum and causing the orbit to decay further \citep{Fla77,Pol94,Pej16,Pej17,Lu23,Schu24}. 

The immediate structural impacts of this merger scenario are well-established. The merger product will become a star with a helium core that is undermassive compared to the total mass of the star and evolve bluewards in the Hertzsprung-Russell diagram (HRD) \citep{Pod10,Gle13,Van13,Jus14}. \citet{Jus14} evolved a suite of early, high-mass, case B mergers to core-collapse and found that the compactness parameter in some models \citep{Oco11} was high, indicating a surprising density profile outside of the core, though not all showed this behavior. Still, the long-term structural impact of both early case B mergers themselves as well as the modelling assumptions made when constructing the merger product, are only just starting to be explored in detail \citep{Sch24}. 

There is presently not a consensus on how to model the mergers or their subsequent evolution. Large suites of merger and evolution models are prohibitive in 3D, which requires us to turn to 1D evolutionary models. One workaround is to simulate the merger in 3D and then continue the evolution in 1D, as in \citet{Sch19},\citet{Cha20}, and \citet{Shi24}, but even this is computationally expensive and unfeasible for suites of merger models. To compute a grid of models, only doable in 1D, one must gloss over the details of the merger itself and instead approximate the structure of the merger product. \citet{Jus14} and \citet{Sch24} did this by rapidly accreting onto the primary a mixture of material with the equivalent composition as the surface of the primary but with the total mass of the secondary. \citet{Men17} and \citet{Men24} uniformly mixed in the secondary throughout the entire envelope down to either the boundary of the helium core or at the maximum depth the secondary could penetrate the core. The former method assumes that the secondary is unevolved and will not penetrate the star more deeply than the envelope. The latter method assumes homogeneous mixing.

There is another method to create a merger product presented initially to model the formation of blue stragglers via chance collisions in star clusters. \citet{Lom02} showed that in these dynamical timescale mergers, the merger product will be a mixture of the two stars sorted by the entropic variable, a parameter which is closely linked to the entropy; put simply, the core of the secondary sinks closer to the core of the primary whereas the envelope of the secondary mixes with the envelope of the primary. This method of constructing a merger product well approximates the final structure of the merger found in 3D hydrodynamics simulations, though a recent 3D MHD model of a massive main sequence merger did not produce a final structure that matched what would be predicted by entropy sorting \citep{Sch19}. \citet{Ren20} adopted a simplified version of entropy sorting to create a massive stellar merger arising from mass transfer instead of a chance collision. This method is not without its uncertainties. Entropy and entropic variable are not equivalent and in the absence of shock heating, this method underestimates the amount of entropy added to the merger product.

The short- and long-term structural ramifications of generating merger products using different methods are unclear, but we know that the adopted method does matter. \citet{Neu22} found that rapid-accretion, entropy-sorted, and \texttt{Make Me a Massive Star} \citep[a code which includes shock heating, rotation, hydrodynamic mixing, and mass ejection][]{Gab08} models of a massive case A merger all evolve differently in an HRD and have different composition profiles at key evolutionary states, like hydrogen and helium exhaustion. Given the importance of massive stars in shaping the observable Universe and the high rate of mergers of massive binaries, it is critical to understand the evolutionary impacts of these complicated interaction histories. 

In this paper we present a merger model consisting of an 11 and 6.6 \Msun~binary which merges while the primary is crossing the Hertzsprung gap. We selected this model because merging is a common pathway for primary stars of this mass. Approximately 12\% of binaries containing an 11 \Msun~primary will experience an early case B merger, based on the binary initial conditions in \citet{Moe17}, the Milky Way's star formation rate \citep{Lic15}, and a Salpeter initial mass function \citep{Sal55}. We outline our motivation for this test case as well as our modeling choices in Section 2 and compare the merged star structure at carbon ignition to various test models in Section 3. We discuss the usefulness and limitations of the entropy-sorted model in Section 4, and conclude in Section 5. 

\section{Stellar models}

We adopt as our fiducial model a binary system with an 11-day circular orbit, an 11 \Msun~primary, and a 6.6 \Msun~secondary. At this initial period, mass transfer will not initiate until the primary has evolved off the main sequence and into the Hertzsprung gap. The mass of the primary was selected to be massive enough to reach core-collapse without facing numerical problems caused by carbon ignition in a degenerate core. The mass ratio of 0.6 is consistent with the range suggested for early case B mergers which occur as a consequence of the donor expanding into contact \citep{Pol94,Jus14}. 

We construct and evolve our merger product in five steps based on the method put forth in \citet{Ren20}. The steps, laid out in detail below, include evolving the binary from birth through the onset of mass transfer to the point where the stars come into contact, followed by the manual construction of a merged star and finally, the evolution of the merger product to carbon ignition. This technique is generalizable to any merger that occurs on a timescale more rapid than a thermal time. 

We use version 15140 of the Modules for Experiments in Stellar Astrophysics (\texttt{\texttt{MESA}}) 1D stellar evolution code \citep{Pax11,Pax13,Pax15,Pax18,Pax19,Jermyn2023}. All models are non-rotating and solar metallicity (Z = 0.02). Inlists, saved models, profiles, and history files are available for download at \url{https://zenodo.org/records/15490275}.

\subsection{Initial conditions and evolution}
We evolve our fiducial model using the binary module within \texttt{MESA} \citep{Pax15}. All controls not named below were adopted at their default value in this version of \texttt{MESA}. \texttt{MESA} adopts a variety of different microphysics depending on the properties and evolutionary state of the star. Appendix B describes the microphysics in more detail.

We do not treat the companion as a point mass, instead actively evolving both stars according to the same physical assumptions. We include thermohaline mixing and semi-convection, both with efficiencies of 1, in addition to exponential convective overshooting, adopting the overshooting parameters for both core and shell burning of \citet{Cho16}. $\alpha _{\mathrm{MLT}}$ is set to 2. We use the scheme from \citet{Kip80} to determine the thermohaline diffusion coefficient but note that this assumes very efficient diffusion. We adopt the \texttt{basic.net} nuclear reaction network, which includes eight different species ($^1$H, $^3$He, $^4$He, $^{12}$C, $^{14}$N, $^{16}$O, $^{20}$Ne, and $^{24}$Mg), but also enable \texttt{auto\_extend\_net}, which automatically updates the nuclear reaction network to \texttt{co\_burn.net} when \texttt{basic.net} is no longer sufficient. This switch takes place prior to carbon burning but after helium is ignited. \texttt{co\_burn.net} includes $^{28}$Si in addition to the eight species in \texttt{basic.net}.

We assume conservative mass transfer using the \citet{Kol90} scheme but also include mass loss from winds, using the `Dutch' scheme with a scaling factor of 1.0 for hot and cold winds \citep{Gle09}. This scheme adopts a prescription from \citet{Nug00} or \citet{Vin01} (for hot winds) and \citet{deJ88}, \citet{Nie90}, or \citet{van05} (for cold winds) depending on the star's effective temperature and surface hydrogen mass fraction. We do not assume that the accreting star accretes any material lost in the stellar winds from the primary. We allow the orbital angular momentum to change as a function of mass loss and mass transfer only. Other mechanisms to carry away spin and orbital angular momentum, including magnetic braking and gravitational wave radiation, are negligible, thus we do not include them. 

Mass transfer starts once the primary leaves the main sequence. Over the span of about 5000 years, which is $\sim$ 10\% of the total time for an 11 \Msun~star to cross the Hertzsprung gap, the mass transfer rate rapidly increases, reaching values of order 10$^{-3}$ \Msun\,year$^{-1}$ (see \Fig{mass_loss}). Once the mass transfer rate exceeds 10$^{-4}$ \Msun\,year$^{-1}$, the secondary can not thermally adjust to the new material flowing onto it and it rapidly increases in size, tripling its radius in a few thousand years and ultimately filling its Roche lobe.

We terminate the evolution of the system at contact, where both stars fill their Roche lobes. Although there is uncertainty about whether the contact binary will indeed merge \citep{Pol94,Pod10,Jus14}, the continued expansion of both stars as they evolve coupled with their being in contact suggests that a merger is likely. Furthermore, the initial conditions of our binary put it in a domain where mergers are probable, so the time contact is reached is an adequate stopping condition.

\begin{figure}
    \centering
    \includegraphics[scale=0.52]{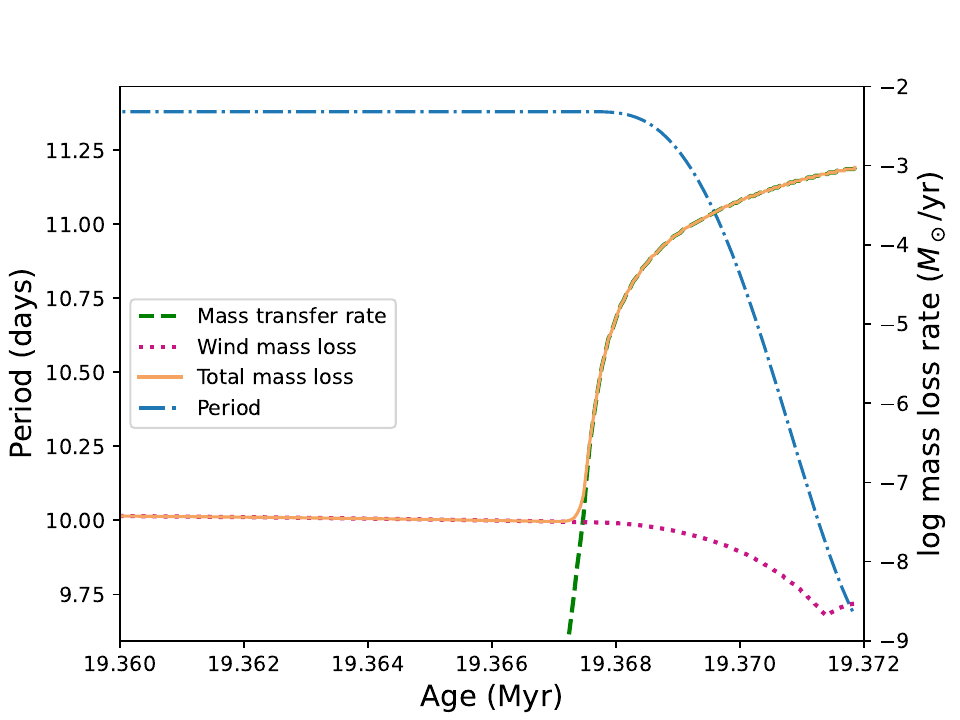}
    \caption{The period evolution and mass transfer/mass loss rate for the primary star in the 11 + 6.6 \Msun~binary for the last $\sim$10 000 years prior to the contact phase. As the mass transfer initiates and rapidly increases, the period shrinks.}
    \lFig{mass_loss}
\end{figure}

\subsection{Simulating the merger}
\texttt{MESA} cannot model the process of two stars merging itself, so we instead have to simulate the effects of the merger on a 1D mass and composition profile. To do this we consider three limiting cases. The first is the rapid-accretion model of \citet{Jus14} and \citet{Sch24}, which are presently the only suites of different merger products evolved to core-collapse. This model assumes that material accreted from secondary onto the primary has the same entropy and composition as the surface of the primary. In this scenario, the mass of the merger product is changed, but the composition of the envelope remains untouched immediately post-merger \footnote{Mergers are subject to additional mixing processes, including hydrodynamical mixing from the impact itself, though this is minimized if the merging stars are different spectral classes \citep{Gab08}, and rotational mixing. We do not consider these extra mixing processes in our models.}. 

Our second case is a modification of the rapid-accretion model. Instead of accreting material to the surface of the primary with the same composition as the envelope of the primary, we accrete material with the average composition of the secondary at the time of contact. This way, chemical enrichment, particularly in helium, from the secondary's main sequence evolution is considered. When this material is accreted, there will be extra thermohaline mixing in the envelope.

Our third case is the entropy-sorted model of \citet{Ren20}, which takes a rapid-accretion model and then relaxes its composition to a mixture of the primary and secondary's compositions ordered by increasing entropy. This method is motivated by the results from \citet{Lom02}, which showed that the final product of 3D hydrodynamics merger models is well-approximated in 1D by a new star whose composition is a mixture of the two stars, sorted by the entropic variable. While not exactly equal to entropy, we can approximate the entropic variable by the specific entropy (entropy per baryon, dependent on the temperature, density, and composition of the star \citet{Cla83}). Entropy-sorting makes some intuitive sense. Stars have enormous temperature and density gradients and will interact on a timescale shorter than that of heating and cooling. It is therefore natural to expect that the hot and dense, lower-entropy material in stellar cores will sink to the center of the potential well whereas the cooler, less dense, higher-entropy material stays towards the surface. 

The rapid-accretion models are constructed in one step. We rapidly accrete the total mass of the secondary onto the primary using the \texttt{relax\_mass} function in \texttt{MESA}, where we assume that no mass is ejected during the merger itself. The combined mass of the stars is 17.3031 \Msun, down 0.3 \Msun~from the initial combined mass of the system because of wind mass loss. This function incrementally increases the mass of the envelope at a fixed rate all while the evolution of the star itself is paused. The composition of this mass can either match the envelope of the star growing in mass or be manually set. We must pick a rate fast enough so that the primary cannot thermally adjust to the mass being added. \citet{Van13} argued that so long as this rate is shorter than the thermal timescale, the properties of the merger product do not vary significantly. We adopt an accretion rate of 10$^{-2}$ \Msun yr$^{-1}$ to match the setup in \citet{Jus14}. 

We require an additional two steps to construct the entropy-sorted model. After the mass is accreted, we then relax the composition of the merger product using the \texttt{relax\_composition} function in \texttt{MESA}. We read in a new composition profile created by sorting the compositions of both stars by their entropy at the time they reach contact, our nominal point at which the merger takes place. We account for composition differences in the eight different species used in our adopted nuclear reaction network. 

\Fig{profs} shows the entropy profiles of the two stars at the time of the merger. Since the primary has overall lower entropy than the secondary, material from the secondary only mixes as deep as its minimum entropy, which is outside the core of the primary. The secondary also has higher maximum entropy than the primary, meaning that part of the secondary floats unmixed at the top of the star. However, there is very little mass with entropy between the maximum entropy of the primary and secondary. Thus, the structure of the merger product is qualitatively the untouched core of the primary embedded in the mixed core-envelope boundary and envelope of the primary and secondary, with a small sliver of purely the secondary's envelope on top, as is shown in \Fig{profs}. These regions are not unique to this particular model; any early case B merger, except from an equal mass binary, should have the same general composition profile. As the mass ratio nears 1, the amount of overlap between the two stars increases.

This starting structure is different from the rapid accretion models because the core-envelope boundary and the base of the envelope of the primary is disrupted by the mixing in of the secondary due to the injection of material deeper into the star. In the rapid-accretion scenarios, the secondary mixes into the envelope either at the surface or more deeply via thermohaline mixing if the composition differs from the envelope of the accretor. Studies which accrete material with the equivalent composition as a the surface of the primary to the surface of the primary implicitly assume the secondary to be unevolved. Accreting mass with the average composition of the star at the time of the merger captures the chemical enrichment from evolution but not the temperature and density gradients. The entropy-sorting approach considers the full evolution of the secondary, capturing changes in density, temperature, entropy, and chemical abundances, which we see in the new composition profiles. The plateau in the helium profile shown in \Fig{profs} corresponds to the core of the secondary and the subsequent comb-like structure is the mixing of the base of the envelope of the secondary with the core-envelope boundary of the primary. Outside the core-envelope boundary of the primary, the envelopes of both stars mix cleanly together.

Once the merger products have thermally relaxed, the jagged peaks in the entropy-sorted helium profile disappear, leaving instead a region of helium enriched material outside the core spanning about a 9 \Msun. 

\begin{figure*}
    \centering
    \includegraphics[scale=0.52]{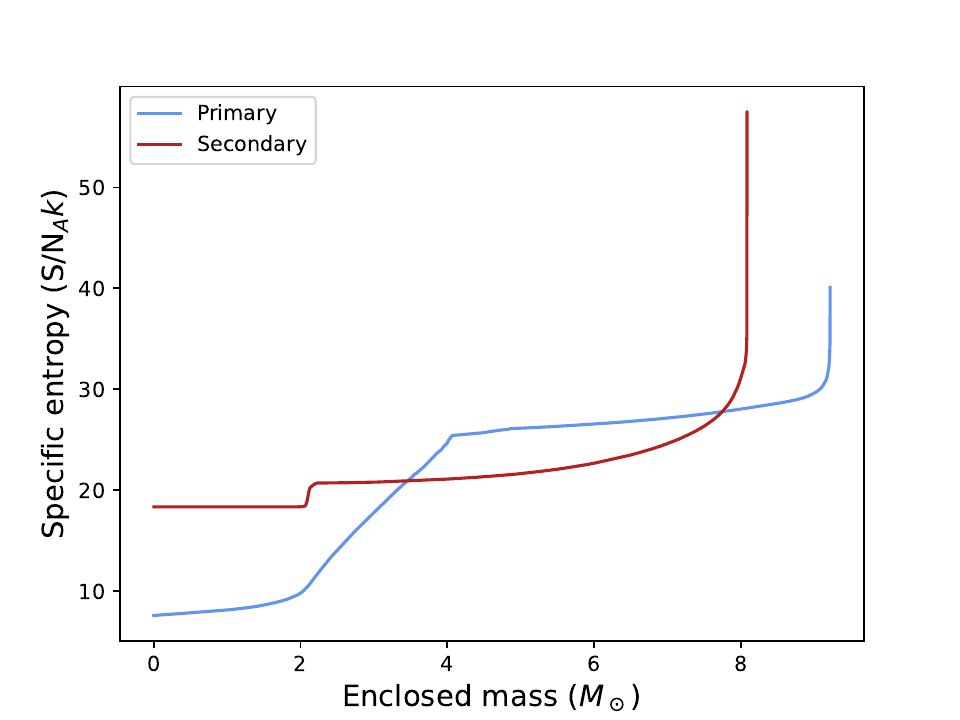}
    \includegraphics[scale=0.52]{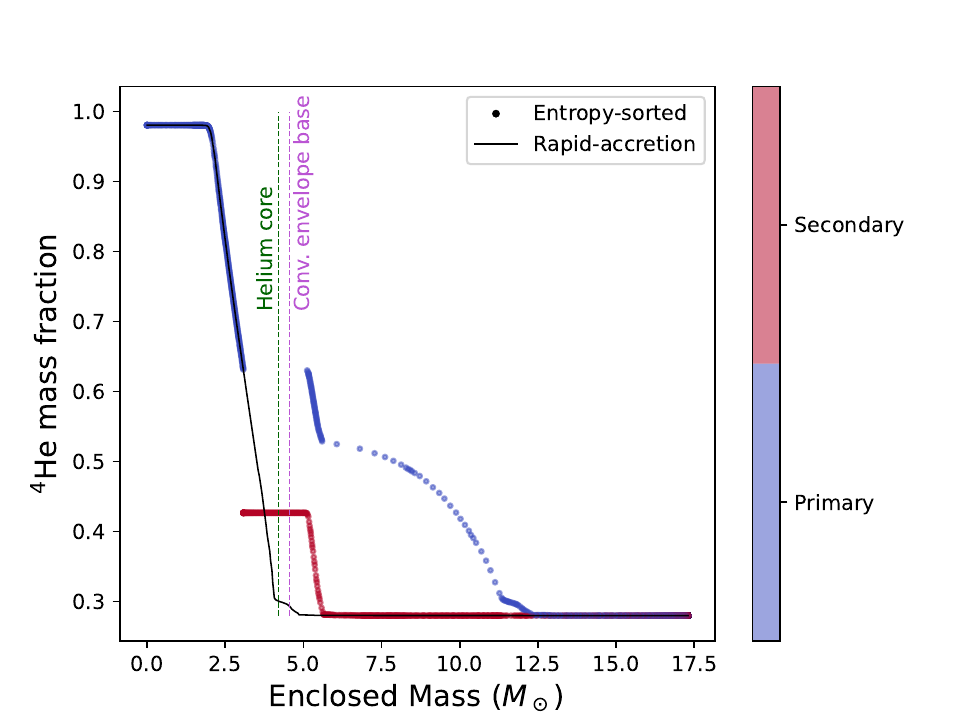}
    \includegraphics[scale=0.52]{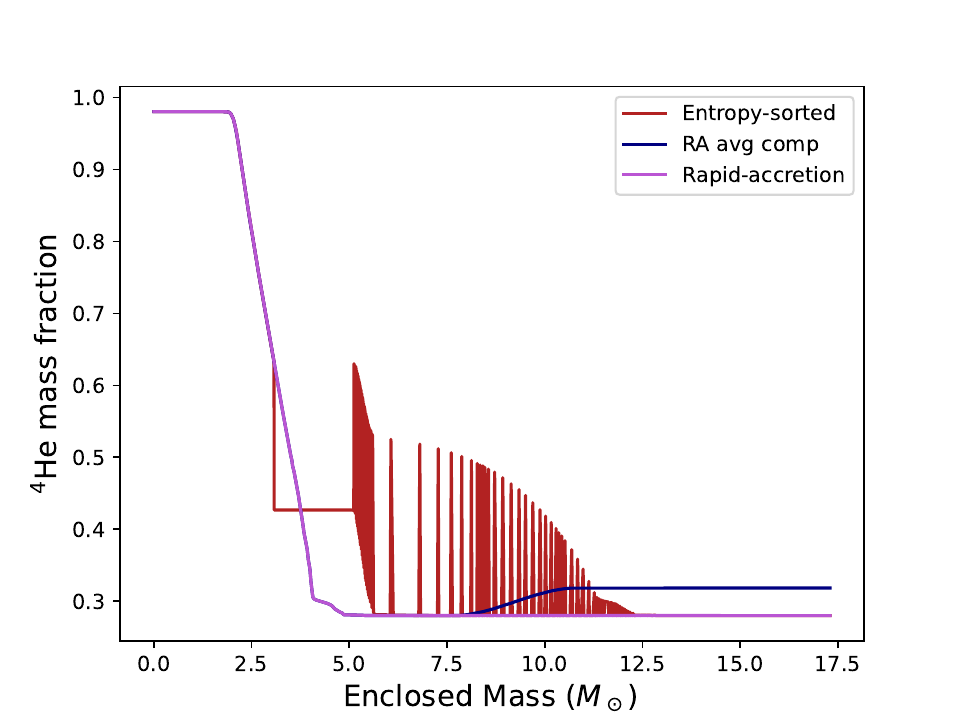}
    \includegraphics[scale=0.52]{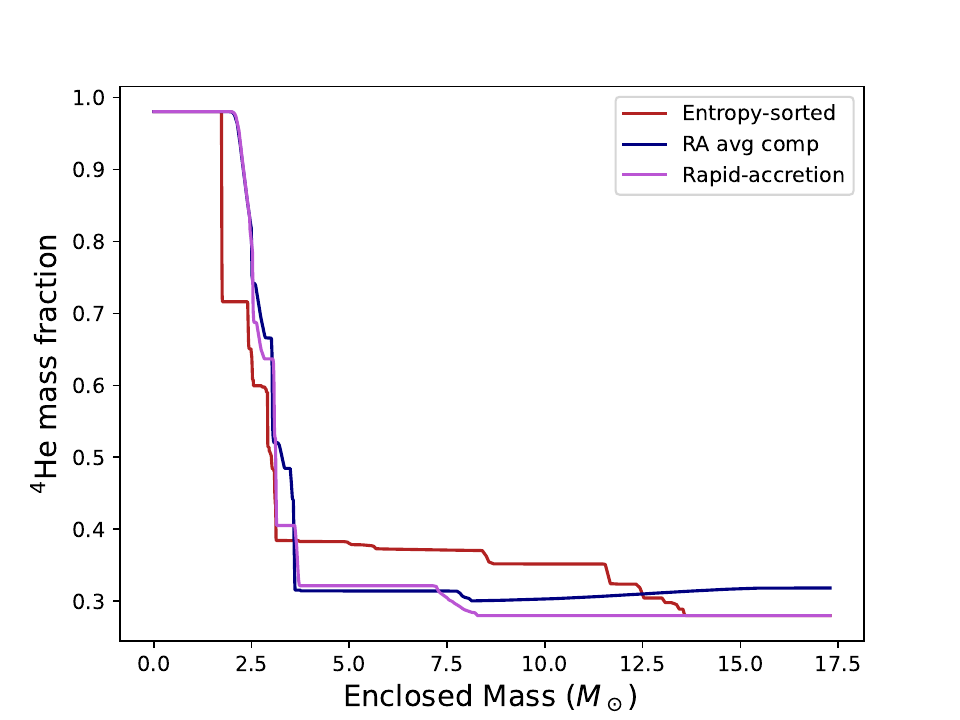}
    \caption{\textit{Top left}: The entropy profiles of both the primary and secondary immediately preceding the merger. The structure of the merger product will be determined by new composition profiles rank ordered by entropy. \textit{Top right}: The entropy-sorted $^4$He profile of the merger product compared to the profiles of the stars individually. The combined profile is colored based on which star contributed the helium in each zone. Vertical dashed lines mark the maximum extent of the helium core and base of the convective envelope at carbon ignition. \textit{Bottom left}: The helium profiles of the entropy-sorted (dark red), rapid-accretion (pink), and composition-averaged rapid-accretion (navy) models immediately post-merger. \textit{Bottom right}: The helium profiles of all three merger models after they have thermally relaxed. The spikes in the entropy-sorted model have settled into a more coherent distribution of helium.}
    \lFig{profs}
\end{figure*}

We deviate from the method put forth by \citet{Ren20} by including a brief period of thermal relaxation between the mass and composition relaxation to let the envelope adjust to the mass accretion. The envelope's thermal time is $\approx$ 1300 years. The thermal structure of the star is set by both the mass of the star and its composition, so after the rapid-accretion step, the star is thrown out of thermal equilibrium and must settle into its new structure. If we were to then change the composition while the star was out of thermal equilibrium, we would thrust it further out of equilibrium, making it difficult to relax and for \texttt{MESA} to treat numerically. \Fig{therm} compares the temperature profiles of the model at various points in the process with and without the thermal relaxation. If we include a thermal relaxation in between when the the mass is rapidly accreted and when the composition changes, then the structure of the star prior to and after the composition relaxation are very similar and \texttt{MESA} can run easily. If we do not include the thermal relaxation, the temperature profile after changing the composition varies by an order of magnitude in some places.

\begin{figure}
    \centering
    \includegraphics[scale=0.52]{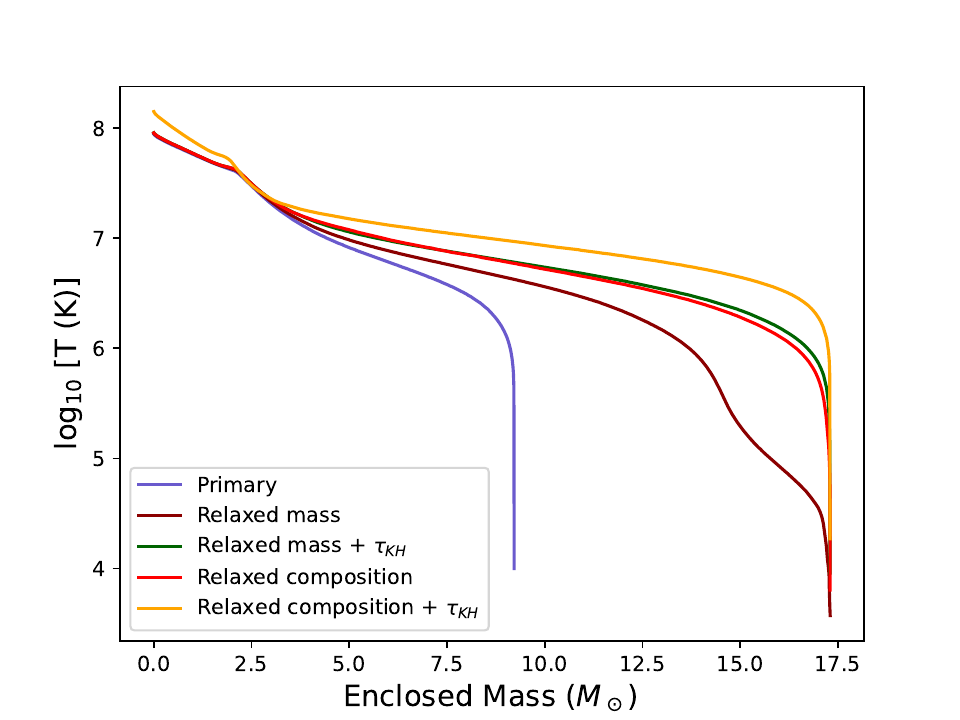}
    \caption{The thermal structure of the merger at the end of each of the first three steps in the creation of the merger product, with (`+ $\tau_{KH}$') and without thermal relaxation of roughly a thermal time (3500 years) in between the mass relaxation and the composition relaxation. The thermal profile of the primary is also shown for comparison. If the thermal relaxation is not included, the temperature gradient is steeper and the temperature outside of the core is an order of magnitude cooler.}
    \lFig{therm}
\end{figure}

\subsection{Post-merger evolution}
Once we relax the composition of the new star, we have created our merger prototype. The final step of this process is to continue the evolution of the merger product. We evolve the star to the point at which its central temperature is $\sim$ 500 million K (log T = 8.7), which is about the temperature at which carbon ignites in the core. We adopt the same mass loss, reaction network, and mixing parameters as prior to the merger.

In summary, the steps taken to get the merger models to carbon ignition  are:
\begin{enumerate}
    \item Evolve the binary to the point of contact.
    \item Rapidly accrete the total mass of the secondary onto the surface of the primary (using \texttt{relax\_mass}).
    \item Evolve the merger product to carbon ignition.
\end{enumerate}
for the rapid-accretion models, and:
\begin{enumerate}
    \item Evolve the binary to the point of contact.
    \item Rapidly accrete the total mass of the secondary onto the surface of the primary.
    \item Relax the post-accretion model over the Kelvin-Helmholtz time of the envelope.
    \item Change the composition of the post-accretion, thermally-relaxed model to that of both stars sorted by entropy (using \texttt{relax\_composition}). 
    \item Evolve the merger product to carbon ignition.
\end{enumerate}

\subsection{Comparison Models}
To demonstrate how the outcomes of such mergers have structures different from those of single stars, we evolve a suite of comparison models that share various properties with the merger product. The first two are single stars with initial masses equal to the primary's initial mass (11 \Msun) and the merger product's mass (17.3031 \Msun)\footnote{For the latter we do use this precise mass for the model, though it is abbreviated to 17.3 for the remainder of this text.} The next test case is a single star with an initial mass of 13.6 \Msun. This model was selected because it has an equivalent helium-core mass at carbon ignition, and as such, an equivalent carbon-oxygen (CO) core mass to the merger product. As has been demonstrated repeatedly, CO-core mass at carbon ignition is a critical parameter for setting the evolutionary track a star will follow until core-collapse. 

The rapid-accretion models described above provide a second source of comparison. Comparison with these models show that using our entropy-sorting scheme leads to significantly different interior structures than mergers approximated as mass gains in the envelope.

\section{The impacts of merging on structure and evolution}

Key diagnostics show the short-term and long-term structural impacts of merging. We already showed that the thermal structure of our entropy-sorted and rapid-accretion models vary significantly post-merger in \Fig{therm}. Now we will explore the long-term differences by looking at probes of the internal structure and evolution in an HRD.

\subsection{Changes to the core structure}

Even without evolving the models all the way to core-collapse, we see structural changes by the onset of central carbon-burning that will effect the long-term evolution of the models. Properties of the CO-core at the onset of central carbon burning will put the star on a unique evolutionary track, barring subsequent interaction, which will determine the star's final structure and final fate \citep{Tim96,Suk14,Pat20,Suk20,Lap25}. The two key properties are CO-core mass and carbon mass fraction \citep{Tim96,Suk14,Pat20,Suk20,Lap25}. 

\Tab{xco} lists the helium-core mass, CO-core mass, and carbon mass fraction at carbon ignition for each model. Core masses are defined as the enclosed mass where the mass fraction of the previous fused species equals 0.01. We find that the helium core mass of our entropy-sorted merger product is undermassive compared to an initially single 17.3 \Msun~model, meaning the merger product was not significantly rejuvenated, an unsurprising outcome since the helium core of the primary was already well developed at the time of the merger. This is consistent with \citet{Gle13}, \citet{Van13}, \citet{Jus14}, and \citet{Sch24}. The core is also more massive than the 11 \Msun~model, meaning the extra mass added in the merger as well as the injection of helium-rich material deeper into the star via entropy-sorting, does affect the duration and extent of hydrogen shell burning such that, by mass, the helium core is comparable to a 13.6 \Msun~star. The helium-core mass of the rapid-accretion models are also larger than that of the single 11 \Msun~star, though not as much as the entropy-sorted model.
\begin{table}
    \centering
    \begin{tabular}{lccc}
         \hline
         &  $M_\mathrm{He}$ & $M_\mathrm{CO}$ & $X_\mathrm{C}$\\
         \hline
        17.3 entropy-sorted & 4.25 & 2.53 & 0.155\\
        17.3 rapid accretion & 3.92 & 2.24 & 0.159\\
        17.3 rapid accretion avg comp & 3.96 & 2.28 & 0.147\\
        17.3 single & 5.86 & 3.94 & 0.269\\
        13.6 single & 4.25 & 2.56 & 0.290\\
        11 single & 3.10 & 1.67 & 0.322\\
        \hline
    \end{tabular}
    \caption{Core properties of all 5 models, with masses given in \Msun.}
    \lTab{xco}
\end{table}

More striking than the undermassive cores are very low values of carbon mass fraction $X_\mathrm{C}$, 0.159, 0.147, and 0.161 for the rapid accretion, average composition rapid-accretion, and entropy-sorted models respectively. Stars normally lie in discrete regions of the $X_\mathrm{C}$ - $M_\mathrm{CO}$ plane \citep{Woo19,Pat20,Sch21,Lap21,Pat22}. More massive stars have higher CO-core masses but also have less carbon and more oxygen. The carbon mass fraction is determined by the competition between the triple-$\alpha$ process and the $^{12}$C ($\alpha, \gamma$) $^{16}$O reaction, and it is sensitive to density, temperature, and the abundance of carbon and helium \citep[e.g.,][]{Bur57,Tak18,Far20,Lap21,Meh22,She23}. Higher mass stars have lower-density, higher-temperature cores that favor carbon destruction over carbon production. This is true whether stars have retained their envelopes or have had their envelopes entirely stripped by mass transfer or winds. 

In the case of envelope stripping prior to helium depletion, the maximum radial extent of convection in the helium core shrinks, leaving behind a gradient of helium and carbon. The relative abundances of helium and carbon change, affecting the 3$\alpha$ and $^{12}$C ($\alpha, \gamma$) $^{16}$O rates and ultimately resulting in a higher amount of carbon for a given $M_\mathrm{CO}$ than a star which retains its envelope \citep[e.g.,][]{Lap21}. An example of these two trends is shown in \Fig{xc-mco}, where we plot the properties of single and case B binary-stripped models from \citet{Sch21}. Our single models lie approximately where the \citet{Sch21} models do, with slight variations that can be traced to differences in the adopted model physics. 

These merger models fall well off the trend lines of stars which retain their envelopes and have their envelopes stripped, meaning they occupy a region of this parameter space unattainable from other evolutionary histories. This is important because a star's location in the $X_\mathrm{C}$ - $M_\mathrm{C}$ plane is strongly correlated with its final structure at core-collapse \citep[e.g.,][]{Pat20}. For instance, the amount of carbon in the core at the time of carbon ignition matters for the long-term evolution because it determines whether carbon burning occurs in a convective or radiative environment, which, due to energy and entropy losses from neutrino emission, will determine the length of the subsequent burning stages \citep{Suk20,Lap25}. \citet{Pat20} showed that for a given $M_\mathrm{CO}$, lower $X_\mathrm{C}$, measured at carbon ignition, tended to correspond to cores at core-collapse with higher compactness and iron core mass, which are more difficult to explode \citep[e.g.,][]{Oco11}. It is unclear whether high compactness at core-collapse is a universal feature of early case B mergers. Neither \citet{Jus14} nor \citet{Sch24} found that their models were consistently more or less likely to explode compared to single stars, when using 1D parameterized explosion criteria. A dense grid of entropy-sorted merger models spanning a wide range of masses and evolved to core-collapse is needed to show whether there are any coherent trends in changes to the final structure and explodability caused by merging. 

To understand why these models have such low central carbon abundances, we have to look at how carbon is being produced and destroyed during helium burning. \Fig{evo} shows the changes in central density $\rho_C$ and central temperature, T$_\mathrm{C}$ as core helium burning progresses and the central helium abundance,$X_\mathrm{He, C}$ drops. All three merger models break from the typical progression of single stars in this plane. The rapid-accretion models evolve prior to helium ignition with cores comparable to that of an 11 \Msun~single star. This is expected since the cores of these models, and their immediate surroundings, remain untouched in the merger. Nevertheless, the addition of mass to the star still changes the core, such that these models burn helium first in a colder and denser environment than the 11 \Msun~model, then progress to hotter and less dense cores for the latter half of helium burning.

The core of the entropy-sorted model immediately deviates from the 11 \Msun~model, becoming the coldest and densest of the three merger models. Like the rapid-accretion models, its core environment also changes, becoming the least dense and hottest of the three, approaching the central core conditions of a 13.6 \Msun~single star. 

For about the first half of core helium burning, when the 3$\alpha$ process dominates helium burning, the merger models have cores that are colder and denser than the single star models. This stunts 3$\alpha$ reactions, resulting in less carbon production. As helium burning progresses and the cores of the merger products get hotter and less dense, the $^{12}$C $(\alpha, \gamma) ^{16}$O reaction rates increases, resulting in more carbon destruction. In \Fig{he-c}, we see that the models diverge as they burn helium, with the merger models reaching their peak in central carbon mass fraction, $X_\mathrm{C, C}$, earlier than the single stars, and burning helium via carbon destruction for longer.

Despite the differences in adopted model physics and the manner in which the merger product is constructed, \citet{Sch24} also find low values of $X_\mathrm{C}$ for some of their case B mergers, suggesting that low carbon mass fraction may be common among case B mergers. Ultimately, entropy-sorted merger models across a wider range of masses, mass ratios, and periods will help determine whether this trend is unique to one set of modeling assumptions and present across different initial conditions.

By helium depletion, the entropy-sorted merger model's evolution in central density and temperature is identical to the 13.6 \Msun~single star. However, we cannot assume that its late-stage evolution will mimic that of a 13.6 \Msun~star because of its significantly lower core carbon content. The rapid-accretion models' progression in this plane are identical to one another for the duration of their post-merger evolution. Prior to helium ignition, both models follow the same trend as the 11 \Msun~star but then diverge from the single star model by helium ignition. The rapid-accretion models complete helium burning hotter and less dense than the 11 \Msun~model but colder and denser than the entropy-sorted model.

\begin{figure}
    \centering
    \includegraphics[scale=0.52]{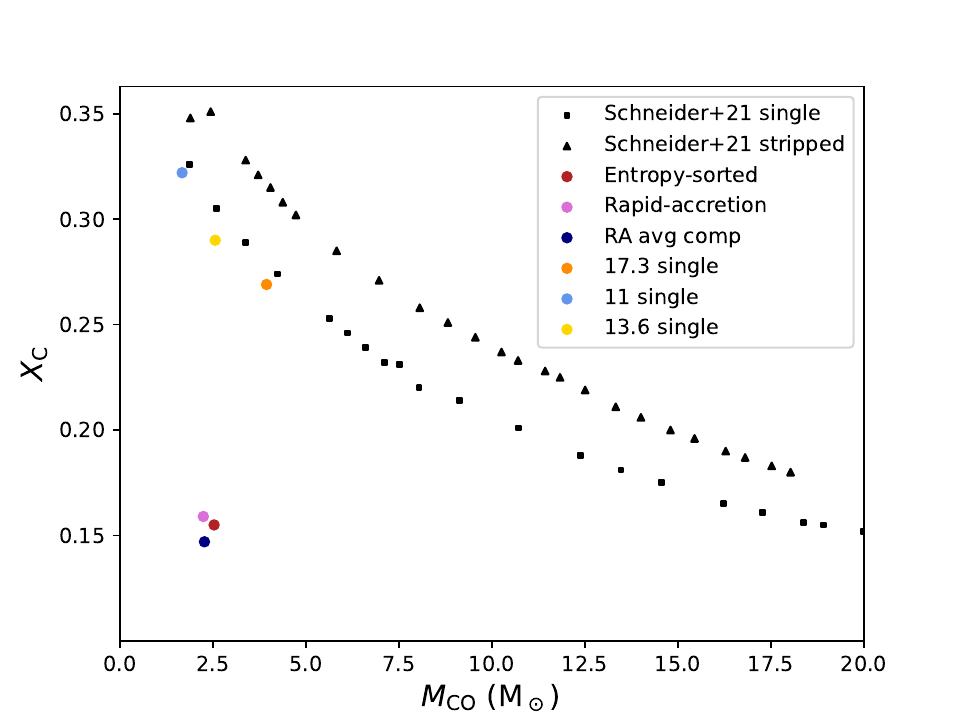}
    \caption{The distribution of $X_\mathrm{C}$ and $M_\mathrm{CO}$ of our models compared to \texttt{\texttt{MESA}} models of single and stripped stars from \citet{Sch21}. Both the rapid accretion and the entropy sorted merger models have significantly lower $X_\mathrm{C}$ than the single and stripped stars.}
    \lFig{xc-mco}
\end{figure}

\begin{figure}
    \centering
    \includegraphics[scale=0.52]{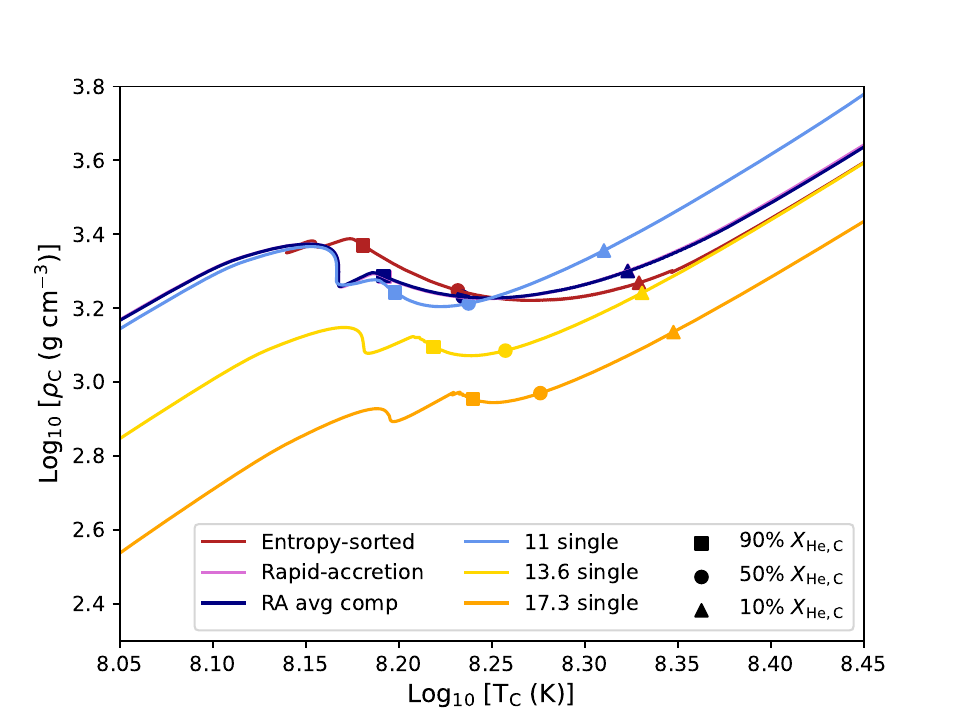}
    \caption{Evolution in central density $\rho_\mathrm{C}$ and temperature T$_\mathrm{C}$ of the merger and single star models. Central helium abundances of 0.9, 0.5, and 0.1 are denoted by squares, circles, and triangles respectively. The rapid-accretion merger models overlap.}
    \lFig{evo}
\end{figure}

\begin{figure}
    \centering
    \includegraphics[scale=0.52]{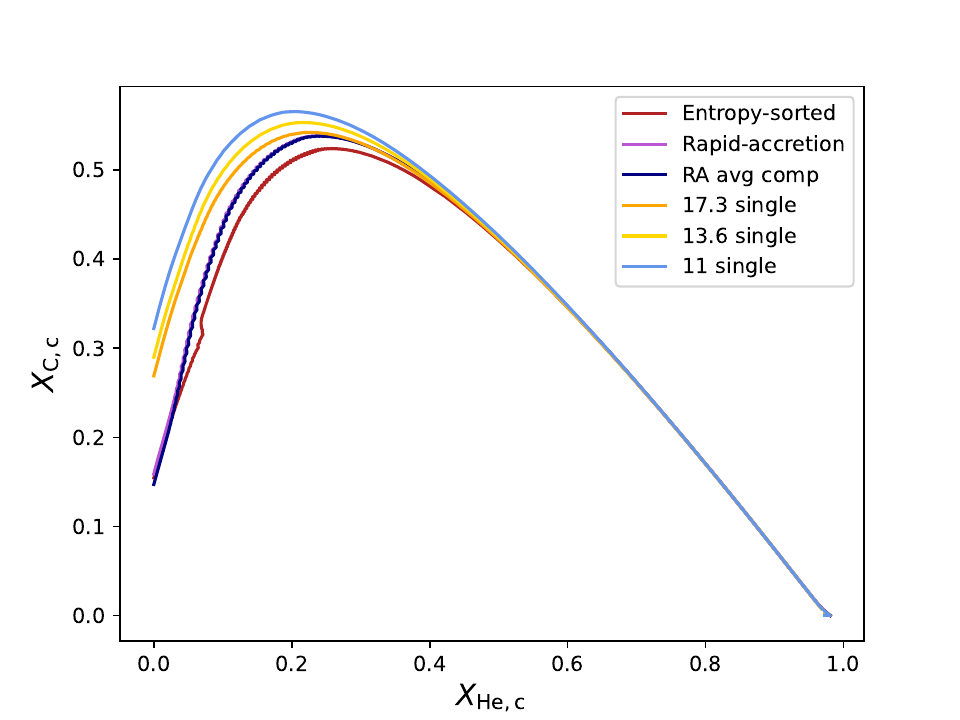}
    \caption{The evolution of both the central $^4$He and $^{12}$C abundance during core helium burning in the different models. Stars move left along the x-axis as time goes on. The core evolution of both rapid-accretion models (navy and purple) is nearly identical.}
    \lFig{he-c}
\end{figure}

\subsection{Surface abundances}
Both the entropy-sorted model and the average-composition, rapid-accretion model show a $\sim$ 70\% $^{14}$N enhancement, $\sim$ 33\% $^{12}$C depletion, $\sim$ 25\% $^{16}$O depletion, and a $\sim$ 20-25\% enhancement in $^4$He in their envelopes over the single star and base rapid-accretion models at the onset of carbon burning. This is because the entropy-sorted and average-composition rapid-accretion models take into consideration the evolution of the secondary, which will exhibit helium and nitrogen enrichment and carbon and oxygen depletion due to CNO burning on the main sequence. Since the base rapid-accretion model assumes the secondary is unevolved, there are no changes to the surface composition.

Merging mixes this CNO processed material from deep inside the secondary into the envelope of the merger product. Enrichment and depletion of this type has been observed in the LMC and reproduced by merger models \citet{Men24}. The log$_{10}$ ${^14}$N/$^{12}$C and ${^14}$N/$^{16}$O ratios for our entropy-sorted model (0.55 and -0.07 respectively) fall well within the bounds of where their ``Group 3" blue supergiants and merger models lie \citep[see Figure 2 in][]{Men24}, though our models started at solar metallicity and theirs at the metallicity of the LMC.

\subsection{Evolution in the HR Diagram}

It is well established that merger products can become blue supergiants \citep[e.g.,][]{Pod92a,Van13,Jus14,Men17,Ren20,Men24,Sch24, Bel24}. As a check of our models we show their evolution in the HRD in \Fig{HRD}. We cannot resolve the short-term behavior of our model before it thermally relaxes, because we are approximating the structure of the merger product without a detailed model of the merger itself. We can only compare the rapid accretion and entropy-sorted models after a thermal time. However, it is interesting to note that the starting structure immediately post-merger in both models is entirely different. As was shown in \citet{Jus14}, the rapid-accretion models become red supergiants immediately post-merger since high entropy material is being added at the surface. Even when the accreted material is helium-enriched, the model immediately post-merger is a red supergiant. It then contracts over a thermal time to a blue supergiant. The entropy-sorted model is born as a blue supergiant because mass is injected deeper into the star, though it also has to thermally relax. In doing so it becomes even hotter. 

In between when the star has thermally relaxed and when helium ignites, there is a brief (order 10$^4$ years) period of peculiar behavior caused by inconsistent hydrogen shell burning. The power from hydrogen shell burning varies by a factor of 5, going through two cycles of increasing and decreasing before increasing while helium ignites. The entropy-sorted merger product completes a loop spanning about a 5000 K change in effective temperature and and 0.1 dex change in luminosity. The rapid-accretion models complete a set of three smaller loops spanning about a 1000 K change in effective temperature and 0.03 dex in luminosity. Though these loops are small, they are still atypical for stellar evolution.

Once the stars have thermally relaxed they ignite helium as blue supergiants. The rapid-accretion models spend about 2.5 million years crossing the Hertzsprung gap whereas the entropy-sorted model spends about 2.9 million years crossing the gap. All models spend the bulk of this time on the blue side of the gap. As has been noted of merger products before, their significantly prolonged lifetimes as blue supergiants can help explain why more blue supergiants are observed than what is expected from single star evolution\citep[e.g., ][]{Fit90,Bel24}. However, the total lifetime of the merger products are not significantly extended. The entropy-sorted model's lifetime is 22.2 Myr compared to 21.9 Myr for the rapid-accretion models and 21.4 Myr for the 11 \Msun~single star.

The entropy-sorted model is more luminous than the rapid accretion models but at most, only by about 0.3 dex. This is not just due to a difference in opacity caused by helium enrichment in the envelope, since the average composition model and the entropy-sorted model display different behavior as they cross the Hertzsprung gap. All models are red supergiants by the time carbon ignites. 

Based on the evolution of our merger products in an HRD alone, it would be difficult to distinguish them from stars that were born single. However, with asteroseismology, it could be possible to detect the undermassive core \citep{Rui21}. Case B mergers may also have a unique pattern of asteroseismic period spacings, which could be detected with long baseline photometry \citep{Bel24}.

\begin{figure}
    \centering
    \includegraphics[scale=0.54]{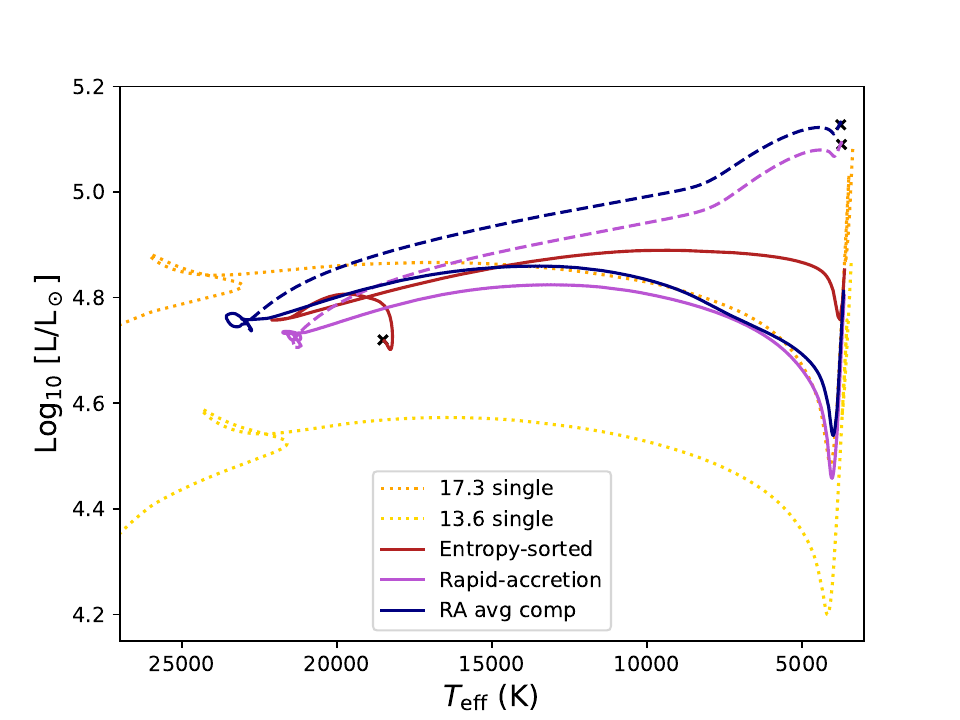}
    \caption{An HRD comparing the post-merger evolution of the entropy-sorted model (dark red) with both the original rapid-accretion model (pink) and the averaged-composition, rapid-accretion model (dark blue). Black crosses denote the start of the post-merger evolution of each model. Dashed lines indicate the time spent thermally relaxing and solid lines show the subsequent evolution.}
    \lFig{HRD}
\end{figure}

\section{Discussion}
We present this merger model as the most simple test case, solely to highlight that a case B merger produces a structure that a single or stripped star cannot produce, and to show that the manner in which the merger product is created, either via rapid-accretion or entropy-sorting, also produces structural differences. That we recover the same general results as \citet{Jus14} and \citet{Sch24} with two different merger setups without attempting to reproduce their models lends credence to the findings. 

Despite the increased complexity of the entropy-sorted model over the rapid-accretion model, entropy-sorting is warranted for higher mass ratio mergers. In these scenarios, the secondary is more evolved, thus the entropy in its core is lower. When the stars merge, the secondary will penetrate deeper into the primary, injecting material with greater helium-enrichment closer to the core. Since the final mass of the merger is higher for higher mass ratio binaries, we expect the helium core mass at carbon ignition to be larger than that of a lower mass ratio merger, which means more of the helium-enriched material from the secondary will be engulfed by the core, enhancing the differences between a rapid-accretion and entropy-sorted merger. 

An example of this is shown in \Fig{He_comp}, which compares the entropy-sorted $^4$He profile of our fiducial merger with the entropy-sorted profile of from an 11 \Msun~and 8.8 \Msun~binary on an 11-day orbit if it were to merge. We offer this qualification because the upper limit in mass ratio for which stars merger via the early case B contact channel is uncertain \citep[e.g.][]{Jus14}. We indeed see that the 8.8 \Msun~secondary interrupts the core-envelope boundary of the primary closer to the core with material that is more helium-enriched than the 6.6 \Msun~primary. 

\begin{figure}
    \centering
    \includegraphics[scale=0.54]{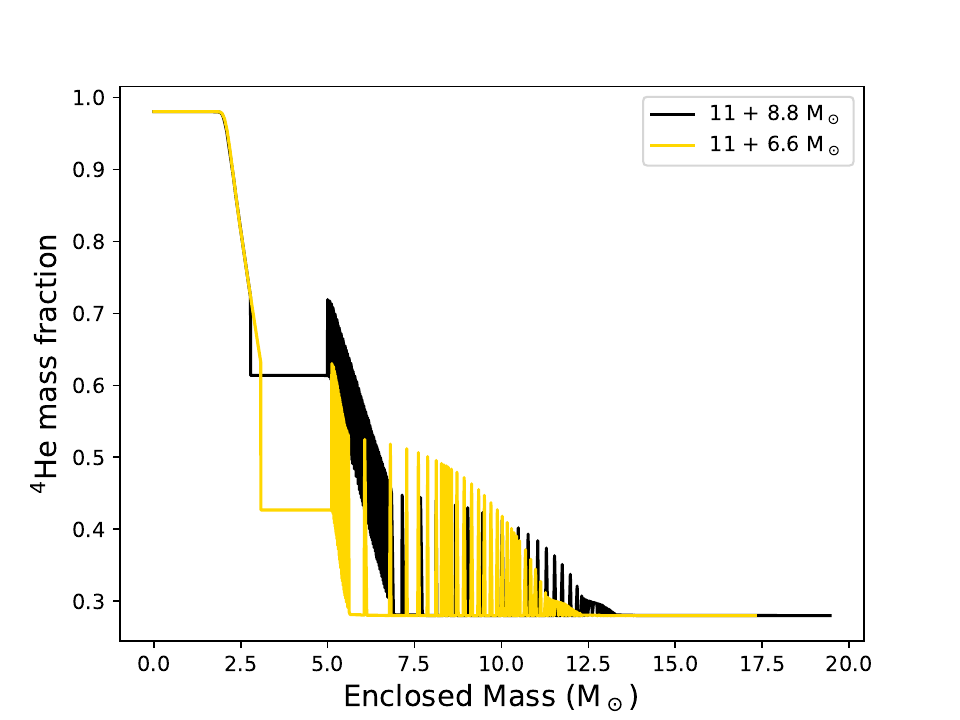}
    \caption{A comparison of the entropy-sorted $^4$He profile from a merger of a binary containing 11 \Msun~and 8.8 \Msun~stars on an 11-day orbit (black) and that of our fiducial model (gold).}
    \lFig{He_comp}
\end{figure}

Still, our models are subject to a variety of uncertainties, from stellar evolution, binary evolution, and the difficulties of reproducing an inherently 3D merger process in 1D. We assume that mass transfer is completely conservative, however, \citet{Hen24} finds that the landscape of models which come into contact completely changes depending on how conservative mass transfer is. We also assume, for simplicity, that the stars are non-rotating and do not spin-up from accretion. However, stars do spin up from accretion and can quickly reach breakup velocities without mediating factors like rotationally-enhanced winds or interaction with a forming accretion disk \citep{Pac81}. The stars in our system merge quickly, however, which limits the amount of time spent rapidly rotating due to accretion. The final rotational velocity of the star post-merger is unclear. \citet{Sch19}'s 3D hydrodynamical model of a main sequence merger found that the merger product rotated slowly, contrary to what one might expect. However, a case B merger has been invoked to explain Betelgeuse's (potential, see \citet{Ma24}) rapid rotation \citep{Cha20,Shi24}. Because of the more extended, convective envelope of the primary at the time of merger, it is easier to sustain a higher rotational velocity \citep{Cha20}. It is possible that whether a merger product is spun-up and maintains its enhanced rotation is a function of the evolutionary state of the stars in the binary.

Ultimately, the final rotational velocity of the merger product, and the amount of time it stays rapidly rotating, can affect the carbon abundance due to rotational mixing and any structural changes caused by rotation-induced oblateness. \citet{Lim18} showed that the distribution of stars in the $X_\mathrm{C}$ - $M_\mathrm{CO}$ plane flattens when stars experience an extended period of rapid rotation. 

Mass loss from ejecta and winds is also uncertain. Some amount of material is ejected from the envelope as the stars merge, which we do not presently account for. \citet{Gab08}, \citet{Ric12}, \citet{Gle13}, and \citet{Bal23} estimate that at most 10-12\% of the envelope is removed in direct collisions. This percentage can vary depending on how bound the envelope is and the impact velocity; the impact velocity during a merger caused by binary interaction may be lower than that of a direct collision. While stars are centrally concentrated, removing mass can still change the thermal structure of the star. Furthermore, the wind mass-loss rates of red and blue supergiants are contested, with some arguing that the schemes presently implemented in \texttt{\texttt{MESA}} overestimate the amount of mass lost \citep[e.g., ][]{Ful06,Smi14,Rom24,Dec24}. 

Metallicity will impact these results too. Not only is mass-loss metallicity-dependent but so are the binary initial conditions over which you get case A, B, and C mass transfer due to changes in stellar expansion \citep{deM08}. Changes in metallicity, rotation, and mass ejection will likely only exacerbate the differences between the structure of a merger product and single stars, and will be considered in a future work. 

\section{Conclusions}

Massive stars play a critical role in shaping the observable universe through their ionizing photons, strong winds, violent deaths, and compact object remnants. The properties of all of these things depend on the structure of the star throughout its lifetime, which, in turn, depends on the star's evolutionary history. Since essentially all massive stars have at least one stellar companion, massive stars are able to frequently take evolutionary pathways inaccessible to single stars or wide binaries.

Merging with a companion, especially as the primary begins to evolve off the main sequence, is a common outcome for massive stars. Therefore, understanding how merger products evolve is essential for understanding massive star evolution as a whole. In this paper, we put forth a test case to show both that early case B mergers can create internal structures unattainable through single star evolution and that the manner in which the merger product is constructed, either via entropy-sorting or rapid accretion, matters for its long-term evolution.

All merger models presented in this work, regardless of the manner in which they were constructed, bear three similar qualities. The first is that they have undermassive helium cores, a property long-known to occur in case B mergers \citep{Pol94,Pod10,Gle13,Jus14,Sch24}. The entropy-sorted model does have a higher mass helium- and CO-core at carbon ignition due to the injection of helium-rich material from the secondary deep into the star as the primary's helium-core grows from hydrogen shell burning. 

The second is that the helium core during helium burning has a unique structure, leading to properties in the core by the onset of carbon burning that are unattainable through other evolutionary histories. All merger models have cores that are cooler and denser at the start of helium burning, leading to less carbon production. The entropy-sorted model has the coolest and densest core of the three merger models. These cores get hotter and less dense during helium burning which leads to more carbon destruction. The entropy-sorted model gets the hottest and least dense of the three merger models, ultimately following the same evolution in central temperature and density as the 13.6 \Msun~model. The rapid-accretion models follow the same evolution in central density and temperature as each other. All merger models end up with nearly half the central carbon abundance as the single star models.

The third similarity is that all merger products become blue supergiants and spend over two million years crossing the Hertzsprung gap. The entropy-sorted model is born as a blue supergiant whereas the rapid-accretion models are born as red supergiants but become blue supergiants by the time they thermally relax. All merger models end up roughly 200-300 Kelvin hotter than the single star models by the onset of carbon ignition and reach a luminosity within a few hundredths of a dex of that of the 13.6 \Msun ~model. 

Still, entropy-sorting matters for the structure and evolution of the merger product. It takes into consideration the evolution of the secondary in terms of both the composition and the decreasing entropy in its core. While composition changes can be somewhat accounted for by rapidly accreting material with the average composition of the secondary at the time of the merger, this material is confined to the envelope. As such, the average composition model is more luminous than the original rapid-accretion model due to its increased helium abundance, however both models' core evolution is identical. The differences between the core structure of the entropy-sorted model and the rapid-accretion models are indeed due to the entropy sorting.

A grid of entropy-sorted merger models is ultimately needed to compare to present grids of rapid-accretion models and determine whether the behavior exhibited by this entropy-sorted model, namely the most extended Hertzsprung gap crossing time, the higher luminosity, a low central carbon mass fraction, and a core structure most different than that of a single star, are universal features of case B merger models or a feature of this particular model. A grid of models will also reveal any bulk differences in observable properties of the evolved merger products, such as systematic shifts to hotter effective temperatures and enhanced nitrogen and oxygen surface abundances. Such a grid will be presented in a future work. 

\begin{acknowledgments}
RAP thanks Stephen Justham for extensive feedback on the manuscript and many insightful discussions, Mathieu Renzo and Paul Ricker for helpful comments on the methods and science case of this paper, Ylva G\"{o}tberg for the entropy sorting script, Ebraheem Farag for assistance with the models, the Kavli Summer Program in Astrophysics for facilitating collaboration on the project, and Fabian Schneider for providing comparison models. This work was funded by The Ohio State University Presidential Fellowship. TAT was supported in part by NASA grant 80NSSC20K0531. 
\end{acknowledgments}

\software{\texttt{Astropy} \citep{Ast13, Ast18, Ast22}, \texttt{Matplotlib} \citep{Hun07}, \texttt{NumPy} \citep{Har20}, \texttt{PyMesaReader} \citep{Wol17}}


\bibliography{ms}{}
\bibliographystyle{aasjournal}

\appendix 
\section{The effect of resolution}

We run a convergence test to verify that our results are not strongly sensitive to changes in spatial and temporal resolution. Resolution was changed for the entire evolution of the system, pre-, during, and post-merger. The models presented in the main body of the text adopt the default resolution of \texttt{MESA}: \texttt{varcontrol\_target} = 10$^{-4}$, which sets the target change in stellar structure with each timestep, \texttt{mesh\_delta\_coeff} = 1.0, which sets the spatial resolution, and \texttt{time\_delta\_coeff} = 1.0, which sets the temporal resolution. We rerun all three merger models at $\pm$33\% spatial and temporal resolution. \texttt{varcontrol\_target} remained at 10$^{-4}$, except in the post-merger evolution for the low-resolution entropy-sorted and average-composition, rapid-accretion models, where we had to increase the value to 10$^{-3}$ in order for these models to run. 

In order for the high-resolution, entropy-sorted models to run post-merger, we had to relax the tolerances from the most restrictive (\texttt{gold2}) to slightly less restrictive (\texttt{gold}). We also had to let the models take the first four steps using the least restrictive default tolerances\footnote{Documentation for tolerances can be found at \url{https://docs.mesastar.org/en/r15140/reference/controls.html\#solver-controls}}. This was due to transient variations in the luminosity profile that spanned orders of magnitude from zone to zone within the star. The first four time steps were on the order of 10$^{-1}$ years, so an insignificant amount of the merger product's evolution was calculated under loose tolerances. 

\Fig{res} shows the values of $X_\mathrm{C}$ and $M_\mathrm{CO}$ for each model at carbon ignition. The rapid accretion are both well converged, showing no variations in $M_\mathrm{CO}$ and little variation in $X_\mathrm{C}$. Because the entropy-sorted models display greater scatter, we run additional models at $\pm$25\% spatial and temporal resolution. Excluding the lowest resolution entropy-sorted model, the final values of $X_\mathrm{C}$ and $M_\mathrm{CO}$ cluster around values of 0.16 and 2.52 respectively with variations of 1.2\% in core mass and 13\% in $X_\mathrm{C}$ with respect to the highest resolution model.

The lack of convergence on single values of $X_\mathrm{C}$ and $M_\mathrm{CO}$ in the entropy-sorted models is unsurprising due to how sensitive both properties are to conditions in the core. \texttt{MESA} is trying to bring a star thrust out of hydrostatic equilibrium back into hydrostatic equilibrium, and the best-fit solution at each timestep for models with slightly different initial conditions immediately post-merger will not be identical. 

Despite variations in the final values of $X_\mathrm{C}$ and $M_\mathrm{CO}$ among the entropy-sorted models, the trends reported in the main body of the text still are present. All merger models form BSGs which slowly cross the Hertzsprung gap and have low values of $M_\mathrm{He}$, $M_\mathrm{CO}$, and $X_\mathrm{C}$ at carbon ignition, and the entropy-sorted models have more massive CO-cores than the rapid-accretion modes.

\begin{figure}
    \centering
    \includegraphics[scale=0.54]{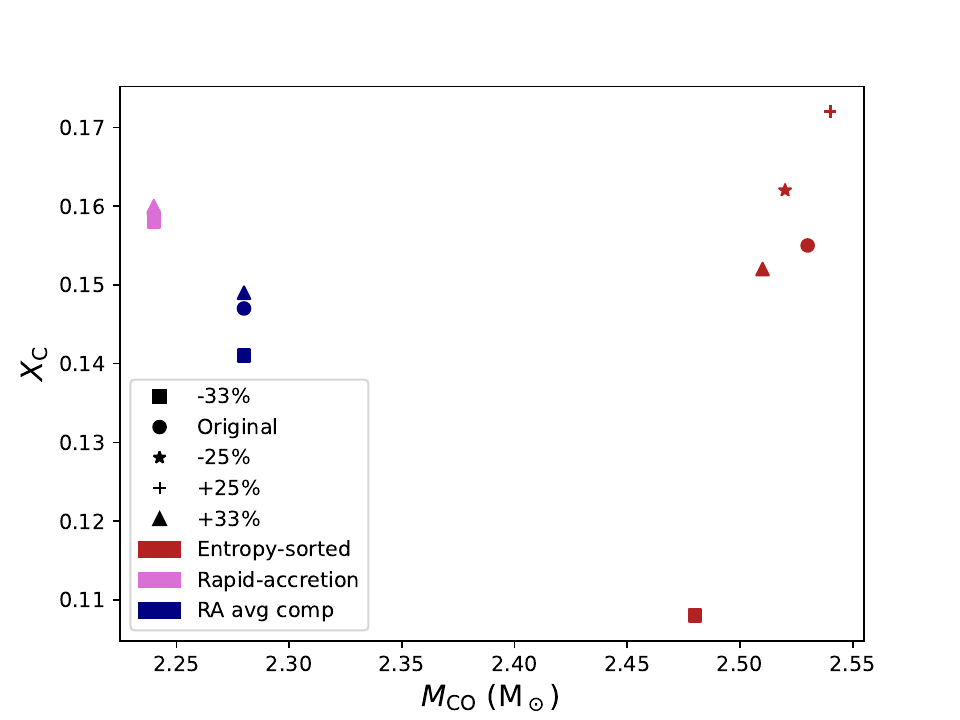}
    \caption{$X_\mathrm{C}$ versus $M_\mathrm{CO}$ for merger models of different spatial and temporal resolution (different marker types). Colors denote the type of merger model.}
    \lFig{res}
\end{figure}

{\section{Adopted microphysics}}
As stated in the main body of the text, we employ the default microphysics unless where otherwise stated. A large breadth of work went into each of these pieces of physics, which we acknowledge from the \texttt{MESA} documentation below:

 ``The \texttt{MESA} EOS is a blend of the OPAL \citep{Rogers2002}, SCVH
\citep{Saumon1995}, FreeEOS \citep{Irwin2004}, HELM \citep{Timmes2000},
PC \citep{Potekhin2010}, and Skye \citep{Jermyn2021} EOSes.

Radiative opacities are primarily from OPAL \citep{Iglesias1993,
Iglesias1996}, with low-temperature data from \citet{Ferguson2005}
and the high-temperature, Compton-scattering dominated regime by
\citet{Poutanen2017}.  Electron conduction opacities are from
\citet{Cassisi2007} and \citet{Blouin2020}.

Nuclear reaction rates are from JINA REACLIB \citep{Cyburt2010}, NACRE \citep{Angulo1999} and
additional tabulated weak reaction rates \citet{Fuller1985, Oda1994,
Langanke2000}.  Screening is included via the prescription of \citet{Chugunov2007}.
Thermal neutrino loss rates are from \citet{Itoh1996}."
\end{document}